\documentclass{optica-article}

\journal{opticajournal} 

\articletype{Research Article}

\begin{document}

\title{Hybrid-Integrated InGaAs/InP SPAD Arrays for Quantum Communications}

\author{Joseph A. Dolphin\authormark{1,2,*}, Rosemary O. E. Scowen\authormark{1}, Louise M. Wells\authormark{1}, David J. P. Ellis\authormark{3}, Abbie L. Lowe\authormark{3}, Benjamin Ramsay\authormark{3},  J. Iwan Davies\authormark{4}, Andrew J. Shields\authormark{1}, Taofiq K. Paraiso\authormark{1}, R. Mark Stevenson\authormark{1}}

\address{\authormark{1} Toshiba Europe Ltd, 208 Cambridge Science Park Milton Rd,
Cambridge, UK\\
\authormark{2} Department of Engineering, University of Cambridge,
Trumpington St, Cambridge, UK\\
\authormark{3} Cavendish Laboratory, University of Cambridge, J. J. Thomson Avenue, Cambridge, UK\\
\authormark{4} IQE PLC, St Mellons, Cardiff, U.K}

\email{\authormark{*}jad218@cam.ac.uk} 


\begin{abstract*}Photonic integration is a promising route to miniaturise the hardware of quantum key distribution (QKD), yet the monolithic integration of single photon detectors remains a significant challenge. QKD receiver chips integrating superconducting detectors have been demonstrated, but their requirement for cryogenic cooling restricts their practical applications. High-frequency gated single-photon avalanche diodes (SPADs) provide a mature non-cryogenic alternative and their fabrication into compact arrays would enable scalable hybrid integration. However, this faces several challenges related to efficient GHz array gating, inter-pixel crosstalk, and scalable waveguide coupling, which to date remain unaddressed. Here, we overcome the key challenges and develop GHz-gated InGaAs/InP SPAD arrays with performance viable for QKD and negligible inter-pixel crosstalk. We combine the arrays with low-loss silica waveguide chips to produce compact hybrid QKD receivers and perform BB84 protocol experiments, achieving secure key rates over 2 Mbps at short distances and 15 kbps over 100 km of fibre. Our work provides a method for flexible and scalable integration of waveguide-coupled SPADs for quantum information processing applications.
\end{abstract*}

\section{Introduction}

For QKD, the ability to implement the technology with compact and practical hardware remains a key limiting factor in its wider adoption. Over the past four decades, QKD has progressed from laboratory experiments to commercially viable products, though it typically still requires bulky hardware significantly larger than an equivalent classical communications device. A promising approach to address this issue is the use of photonic integrated circuits (PICs), which can reduce both the physical footprint and cost of the photonic layer through scalable fabrication techniques. QKD PICs have been widely studied and demonstrated on a variety of material platforms, including silica \cite{Honjo.2004,Tanaka.2012,Sax2023}, indium phosphide \cite{Sibson.2017,Paraiso.2019,Semenenko.2020,Paraiso.2021}, silicon \cite{Sibson.2017b,Ding.2017,Bunandar.2018,Zhang.2019,Geng.2019,Cao.2020,Wei.2020,Avesani.2021}, silicon nitride \cite{Paraiso.2021,Beutel.2021}, thin-film lithium niobate \cite{Lin2025}, and hybrid-material approaches \cite{Dolphin2023}.

However, a notable exception to this trend is the ongoing difficulty of integrating single-photon detectors into QKD PICs. In almost all "on-chip" QKD demonstrations, the detectors remain external and are often the bulkiest and most expensive parts of the system. Implementations of discrete-variable QKD primarily rely on either single-photon avalanche diodes (SPADs) or superconducting nanowire single-photon detectors (SNSPDs). While SNSPDs offer superior performance, their cryogenic cooling requirements (<5 K) limit their applicability in real-world scenarios \cite{Ceccarelli.2021}. In contrast, SPADs can operate at temperatures achievable with thermoelectric coolers, eliminating the need for cryogenic systems. To date, the only examples of detector-integrated QKD chips have used SNSPDs \cite{Beutel.2021,Zheng.2021}, achieving impressive performance (12 Mbps at 10 dB channel attenuation) yet still requiring cryogenic cooling.

An ideal solution would be a near-infrared (1550 nm) SPAD monolithically integrated into the waveguide of a QKD receiver PIC. However, demonstrations of such integration with performance competitive for QKD have yet to emerge \cite{Ceccarelli.2021}. Germanium SPADs on silicon appear the most promising, but still exhibit dark count rates many times higher than their InGaAs/InP counterparts and require cooling to temperatures as low as 100 K \cite{Martinez.2017,Wang.2023}. Consequently, the most practical path toward a non-cryogenic QKD receiver with integrated detectors appears to be the hybrid integration of InGaAs/InP SPADs \cite{Zhang2023,Alimi2025,Ren2024}.

\begin{figure}
    \centering
    \includegraphics[width=1.0\linewidth]{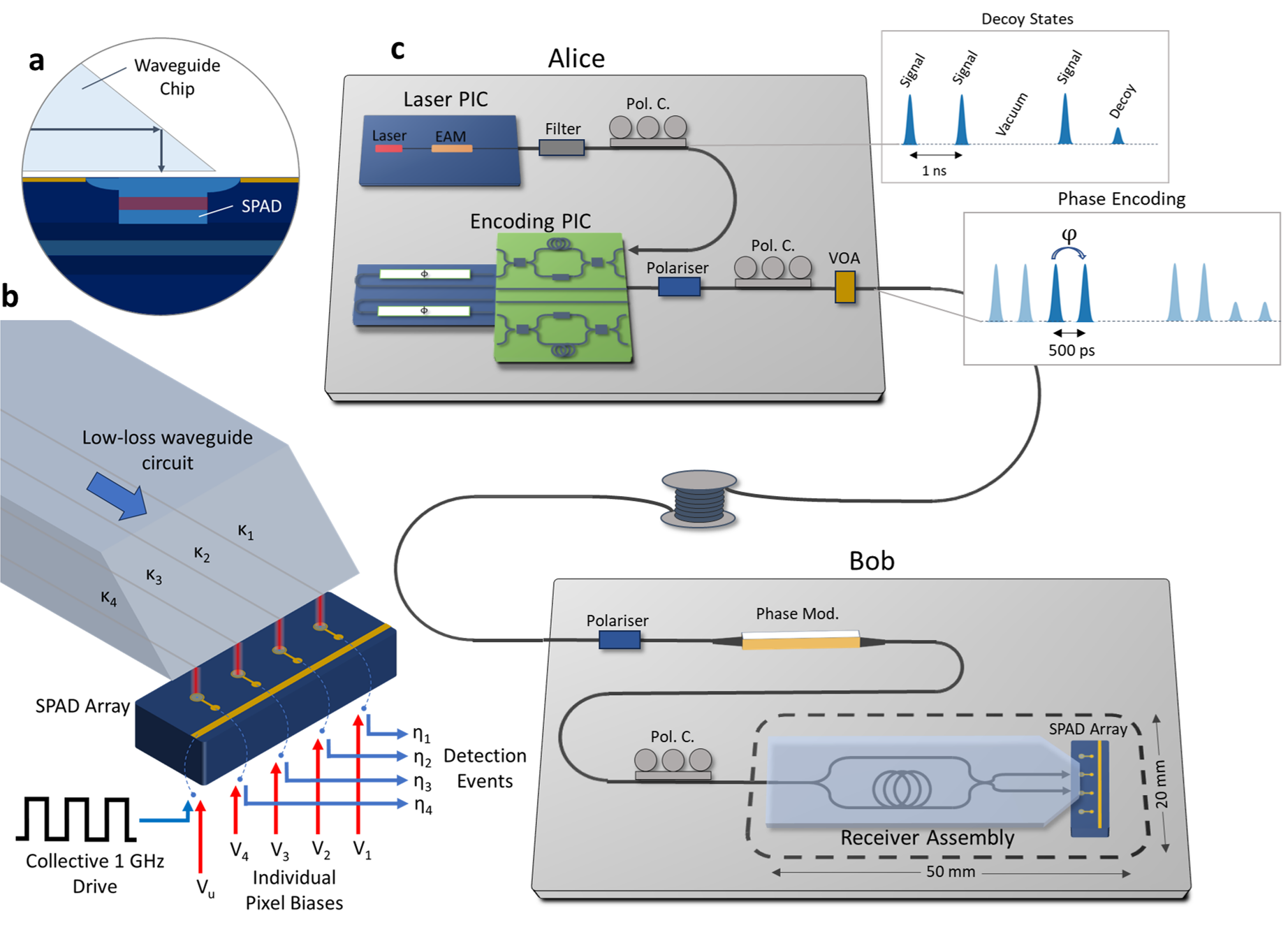}
    \caption{\textbf{a} The quasi-planar coupling technique. A diagonal cut at one end of the silica chip causes light propagating within the waveguides to be reflected downwards and into the aperture of a SPAD. \textbf{b} An illustration of a GHz-gated linear SPAD array with a silica waveguide chip. The entire array is collectively gated at 1 GHz and biased (V\textsubscript{u}), whilst the circuit allows for the DC bias (V\textsubscript{i}) of each pixel to be individually adjusted to compensate for inherent variations in waveguide loss ($\kappa$\textsubscript{i}) and detector efficiency ($\eta$\textsubscript{i}). The pitch of the waveguide array is matched to that of the SPAD array to achieve simultaneous coupling to each pixel. \textbf{c} The photonic circuit for the BB84 QKD experiment. A chip-based quantum transmitter (Alice), is adapted from previous work and is capable of producing three decoy-state time-bin encoded qubits at 1 GHz repetition rate and with low error \cite{Dolphin2023}. An example of the optical signal is shown after the laser PIC and again after the encoding PIC, showing the three-decoy intensity levels and encoding phase $\phi$ respectively. Our quantum receiver (Bob) includes the integrated-SPAD receiver assembly with an asymmetric Mach-Zehnder interferometer.  Polarisation controllers (Pol. C.) ensure polarisation alignment at each stage, a variable optical attenuator (VOA) is used to bring the flux down to secure single photon levels, and a discrete phase modulator provides random basis switching.}
    \label{fig:Concept}
\end{figure}

Here, we report the development of linear arrays of InGaAs/InP SPADs and their application in a hybrid-integrated QKD receiver. Figure \ref{fig:Concept} gives an overview of the work. We use a quasi-planar coupling technique (Figure \ref{fig:Concept}.a) that allows us to efficiently couple the arrays to silica waveguide chips\cite{Han2020}. We demonstrate the first GHz-gating of an InGaAs/InP SPAD array and utilise an electronic circuit that enables efficient collective driving as well as individual pixel bias optimisation to overcome non-uniform SPDE (Figure \ref{fig:Concept}.b). Characterisation of our SPAD devices yields competitive performance: on all pixels a four device array exhibits 15\% detection efficiency, dark count rates per pixel below 8~kHz, and afterpulsing below 4\% at a 100 ns deadtime. For the two devices used in the hybrid QKD receiver we achieve a mean pixel dark count rate of 1.96 kHz, or $1.93\times10^{-6}$ Hz/gate. Further, we find that GHz-gating significantly suppresses the inter-pixel crosstalk. Finally, to demonstrate applicability to quantum communications, we present results from a time-bin encoded BB84 QKD experiment (Figure \ref{fig:Concept}.c). The setup consists of a chip-based transmitter and our hybrid integrated receiver, which combines a low-loss passive silica decoding circuit with an InGaAs/InP SPAD array. The QKD system achieves a positive secure key rate over 100 km of fibre spool and, using room-temperature detectors, we exceed 2 Mbps secure key rate at 0 dB channel attenuation. These results demonstrate a viable path toward compact and practical QKD receivers suitable for metropolitan and access quantum networks.

\section{Implementation}

\subsection{GHz-Gated InGaAs/InP SPAD Arrays}

The InGaAs/InP SPAD arrays were fabricated using a planar geometry based on separate absorption, charge, grading, and multiplication (SACGM) regions \cite{pellegrini2006design}. Optical absorption occurs in the InGaAs layer, exciting a carrier pair that moves through the material under the influence of the electric field, with the hole moving towards the InP multiplication region. Upon entering the InP multiplication region, the hole experiences impact ionisation due to the high electric field, which leads to an avalanche of carrier pairs that is self-sustaining \cite{Jiang2007}. This avalanche of carriers produces a macroscopic current pulse, which can be detected by an external circuit. Grading and charge layers were inserted between the InP and InGaAs: the charge layer tailors the electric field profile, and the grading layer, made of quaternary InGaAsP in step layers, reduces hole trap depth \cite{ramirez2009dependence}. The active area was defined by zinc diffusion through a patterned dielectric mask into the undoped InP layer, forming localized p\textsuperscript{+}-InP multiplication regions. Previous works have described how the quasi-cylindrical junction formed from a single diffusion suffers from premature avalanche breakdown at the edge of this junction, and so a double diffusion process was employed to modify the profile of the diffused junction \cite{ramirez2009dependence, chen2023guard}. Devices were fabricated in linear arrays with a pitch of 250 {\textmu}m.

Our SPADs were operated in gated-mode at 1 GHz with 'self-differencing circuits' used to filter the output signals \cite{yuan2007high}, an approach that has been widely applied to QKD \cite{Yuan2008,Yuan.2018}. However, the above approach has traditionally been applied to single detectors. Figure \ref{fig:Concept}.b shows a conceptual diagram of our supporting circuit; the devices are driven collectively through a shared cathode by the same 1 GHz gate signal, yet retain individual readout channels via separate anodes. Each readout channel used its own self-differencing filter and the DC bias of each pixel could be adjusted individually through a small bias on the anode. Otherwise, no modifications were necessary compared to a conventional self-differencing SPAD circuit. This architecture has three advantages. First, the expensive driving circuit systems (signal generator, amplifier, high voltage DC source) are shared between the detectors, making scaling to larger channel numbers economical. Second, the timing synchronisation of gates between the detectors is implicit in the shared RF drive signal. Third, the adjustability of pixel biases is useful for both individual device optimisation and for QKD security due to the existence of the detector-efficiency mismatch attack \cite{Makarov2006,Fung.2009}.
\begin{figure}
    \centering
    \includegraphics[width=1\linewidth]{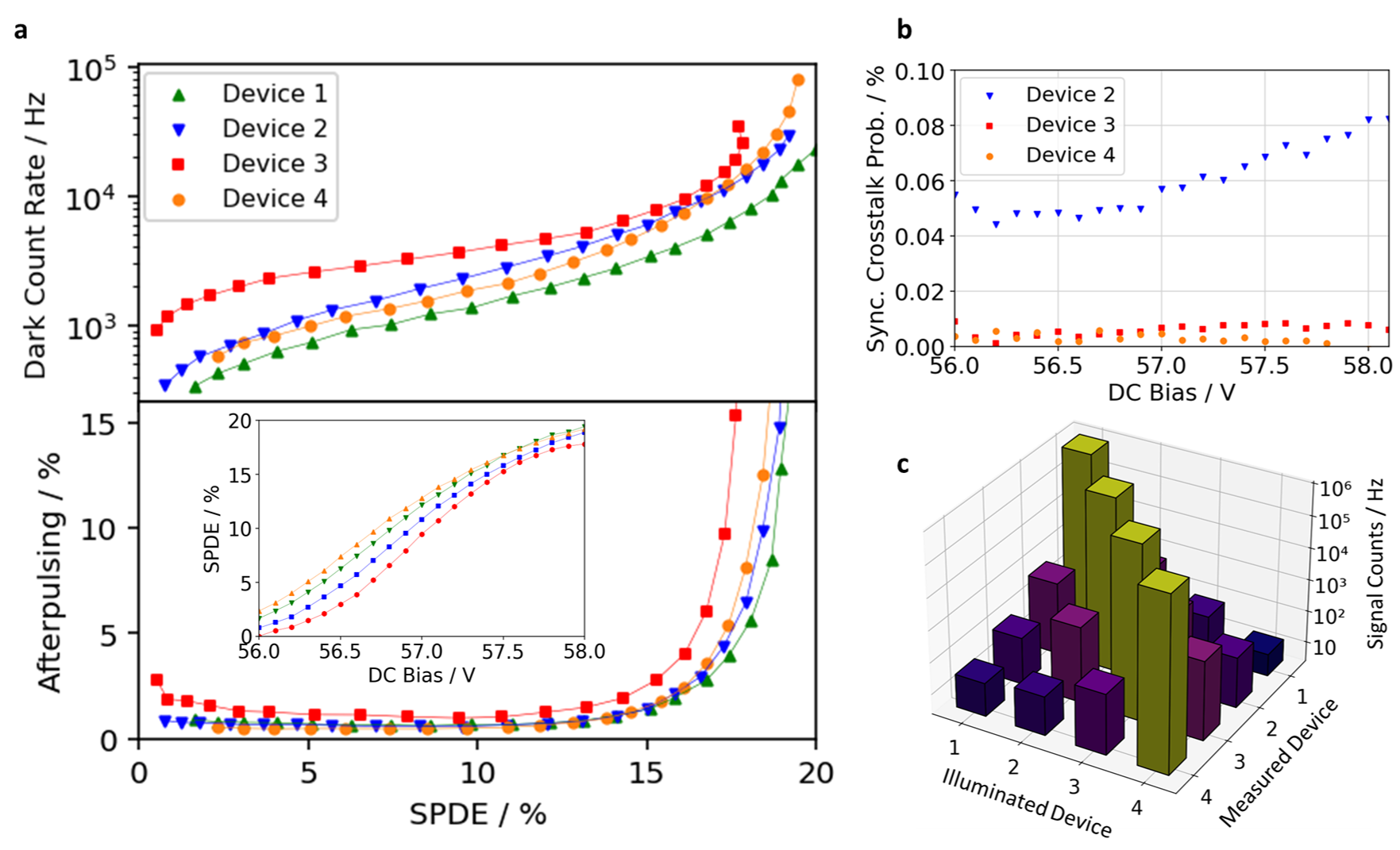}
    \caption{Four-device SPAD array performance during 1 GHz gated-mode operation. \textbf{a} Dark count rate and afterpulsing probability (after 100 ns deadtime) against single photon detection efficiency (SPDE) as the universal array bias is varied. Inset: SPDE against universal array bias across the most relevant voltage range. \textbf{b} Synchronous crosstalk probability across the array for each of the three victim pixels (at successively increasing distance) as the universal array bias was increased. \textbf{c} Array specificity: Detected count rates for the array as each device is illuminated, with dark count rates subtracted. All four devices have their DC bias set to achieve 15\% SPDE. We see specificity of at least three orders of magnitude for all pixel combinations.}
    \label{fig:SPADs}
\end{figure}

The performance of a four-device linear SPAD array was characterised under 1550 nm laser illumination. The driving signal consisted of a 1 GHz square wave, with a 400 ps duty cycle of amplitude 12~V\textsubscript{pp}, and a DC bias set close to the breakdown voltage ($\sim$57 V). A 100~ns deadtime was applied to control the afterpulsing rate. Illumination was with a short ($\sim$30~ps) laser pulse with a repetition rate of 15.625~MHz, such that one in every 64 gates were illuminated. The laser was projected onto the SPAD apertures using a 50x microscope lens with a numerical aperture of 0.42. Each laser pulse was calibrated to contain a mean flux of 0.2~photons, as measured in the plane of the SPAD aperture. Elevated counts in non-illuminated gates were used to calculate afterpulsing probabilities, whilst dark count rates were measured under no illumination. The results of the characterisation are shown in Figure \ref{fig:SPADs}\textbf{a}, demonstrating simultaneous operation across a four-device array with detection efficiencies above 15\%, afterpulsing below 4\%, and dark count rates per device below 8~kHz. The results represent the first reported GHz-gating of an InGaAs/InP SPAD array. 

\subsection{Crosstalk Suppression through Sub-ns Gate Durations}

Whilst compact arrays are advantageous for size and cost, InGaAs/InP SPAD arrays are known to suffer from inter-pixel crosstalk \cite{Itzler2010,Calandri2016,Liu2022,Pesantes.2023,Tang.2024}. The avalanche in one pixel, the `aggressor', can erroneously trigger an avalanche in another pixel, referred to as a `victim'. The probability of this process occurring per avalanche can be referred to as the `crosstalk probability' (see Supplemental Document Section 2.A). For a BB84 QKD system, a crosstalk probability of 1\% increases the quantum bit error rate (QBER) by 0.5\%, as half of the crosstalk events will result in a detection in the incorrect bit state. Some mitigation can be achieved through a universal deadtime for all detectors, such that after a detection event the avalanches in all other detectors are ignored for a set period. In relation to this, we make the distinction between synchronous crosstalk, which occurs in the same gate, and asynchronous crosstalk, which occurs in subsequent gates. Synchronous crosstalk is thus a more serious issue which cannot be filtered out via a universal deadtime. For BB84 QKD systems the theoretical upper bound QBER for a positive secure key rate is 11\%\cite{Amer.2021}, whilst the practical limit for our protocol is closer to 6\% \cite{Lucamarini.2013}. Thus, if we assume that asynchronous crosstalk can be ignored, a 2\% synchronous crosstalk probability is problematic but manageable, whilst a 10\% synchronous crosstalk probability almost completely precludes a positive secure key rate.

Studies of InGaAs arrays have shown that, without mitigation structures, crosstalk probabilities can be as high as 60\% at 240 $\mu$m pixel pitch, falling to 45\% with a metallised trench \cite{Calandri2016}. Another study using isolation trenches reduced crosstalk probabilities to $\sim$1\% \cite{Itzler2009}, dependent on array axis and excess bias. Recent efforts have focused on the complex optical-electronic nature of the crosstalk, creating structures to capture stray electrons and reduce crosstalk by an order of magnitude \cite{Tang.2024}. However, it remains the case that crosstalk is recognised as a serious issue for densely spaced InGaAs SPAD arrays. 

In this work, we find synchronous crosstalk probabilities uniformly below 0.1\% across our GHz-gated arrays, a record-low for equivalent devices by an order of magnitude (see Figure \ref{fig:SPADs}\textbf{b}). Figure \ref{fig:SPADs}\textbf{c} shows the signal count rate across the array during illumination of each pixel, highlighting the specificity of detection events in the array. Though less critical for our application, we find the asynchronous crosstalk probability to be similarly low ($\sim$0.1\%) at typical operating biases. We propose that the cause of this effect is the dynamics of high frequency gating, rather than the structure of the devices. Previous studies in Si and SiC SPAD arrays have identified the \textit{crosstalk formation time}, the time taken for the optical-electronic crosstalk processes to take place, to be on the order of 5-10 ns in these devices\cite{Vila2014,Zhang2019}. We explore this theory in Supplemental Document 2.B. and find an equivalent formation time of 2.5 ns for our devices. Gated operation with shorter duty cycles than this has the effect of cutting off crosstalk before it can take place. This identifies fast gating as a route to reduce crosstalk in InGaAs SPAD arrays, making utilisation for QKD possible and also with wider relevance for other applications.

\subsection{Quasi-Planar Coupling}

Our silica chips were manufactured using an ion-exchange method\cite{Broquin2021,Brusberg2025}, capable of achieving a propagation loss of < 0.2 dB/cm. A mode-field diameter of 10 {\textmu}m was used with a minimum bend radius of 1 mm. To achieve optical coupling between our silica chips and linear SPAD arrays we used a quasi-planar coupling technique, illustrated in Figures \ref{fig:Concept}.a and \ref{fig:Concept}.b. At the output side of the chip an angled cut produces an angled facet. Light propagating horizontally within the waveguides undergoes total internal reflection and is projected downwards. The waveguides run a few microns below the surface of the chip, meaning that the light leaves the chip within a few microns of the tip. The quasi-planar coupling approach was chosen for several advantages. First, it is an established low-loss technique for coupling between silica fibre arrays and PICs. Second, the material silica has low propagation loss which is desirable for a QKD receiver. Third, the coupling orientation allows for the silica chip and SPAD array to be mounted in parallel planes, reducing the height of the final assembly.

With a straight-waveguide chip the quasi-planar coupling approach allowed us to conveniently couple the detectors to individual fibre inputs, whilst for the QKD receiver we could directly functionalise the silica PIC as part of the receiver optical circuit. As the aperture diameter of our SPADs was 20 {\textmu}m, whilst the mode-field diameter was 10 {\textmu}m,  this led to reasonable alignment tolerance in both horizontal and vertical axes. Alignment to arrays was achieved by simultaneously maximising the count rate of the devices at the extreme ends.

In Table \ref{WAFT_coupling} we provide data estimating the coupling loss of the quasi-planar coupling technique. The four-device SPAD array is operated with a balanced SPDE through a four-channel quasi-planar coupling chip, with known channel losses, and then compared to equivalent data with microscope illumination. For the four channels we find estimates of silica-to-SPAD coupling losses ranging from 0.22 dB to 0.68 dB. 

\begin{table}[ht]
\centering
\begin{tabular}{ccccc}
\hline
 & \textbf{Device 1} & \textbf{Device 2} & \textbf{Device 3} & \textbf{Device 4} \\
\hline
\textbf{System SPDE (Silica) / \%}& 10.25 & 10.36 & 10.27 & 10.42 \\
\textbf{Silica Chip Channel Loss / dB}& 1.97& 0.72 & 0.89 & 1.15 \\
\textbf{SPAD SPDE (Microscope) / \%}& 17.0 & 14.3 & 13.8 & 14.3 \\
\hline
\textbf{Silica-to-SPAD Coupling Loss / dB}& 0.22& 0.68& 0.40& 0.22\\
\hline
\end{tabular}
\caption{Estimation of optical coupling loss across a four-device SPAD array using quasi-planar coupling. 'System SPDE' is the efficiency as measured from the input fibre connectors of the silica chip. The individual pixel biases have been adjusted to achieve an approximately uniform system SPDE of 10\%, despite variations in silica chip channel loss. By comparing the system SPDE with the known SPAD SPDE, and after compensating for the silica chip channel loss, we can estimate the optical coupling loss between the silica chip and the SPAD using the quasi-planar coupling technique.}
\label{WAFT_coupling}
\end{table}

\subsection{QKD Receiver Assembly}

Our QKD receiver waveguide chip contained an asymmetric Mach Zehnder interferometer structure with a 500 ps arm length difference and the two outputs of the interferometer proceeding to a quasi-planar coupling angled facet. An optical insertion loss of 1.8 dB  was measured at 1550 nm for the waveguide chip. For the QKD receiver we used two detectors from a three-device array (characterisation in Supplemental Document S2). The three-device array was used due to lower dark counts than the four-device array. The waveguide chip was mechanically held against the SPAD array, rather than permanently attached with glue, to allow us to use both samples for future experiments. The entire assembly was mounted on a cold plate set to  -30{\textdegree}C and with the SPAD array further temperature stabilised to within $\pm$0.01{\textdegree}C by a dynamically controlled thermo-electric cooler. The assembly was held within a nitrogen-purged atmosphere to prevent condensation during cooling. 

The detectors were operated at a SPDE of 15.0\%, mean afterpulsing of 2.23\%, 100 ns deadtime and a combined dark count rate of 3.86 kHz. Individual pixel biases were adjusted to compensate for waveguide chip imperfections and bring the fractional detector efficiency mismatch to <0.01.

\section{BB84 QKD Experiment}

To reinforce the suitability of the SPAD array for quantum communications we performed a chip-based BB84 QKD experiment (Figure  \ref{fig:Concept}.c ). In the transmitter we utilise a previously published edge-coupled hybrid PIC encoder \cite{Dolphin2023}. Within Alice, a 1 GHz train of gain-switched laser pulses was filtered with passband width 0.1 nm and centered at 1552.8 nm. Polarisers and polarisation controllers were used to control the propagation of polarisation throughout the system. A variable optical attenuator at the output of Alice was used to bring the photon flux down to the single photon regime. The communication channel either consisted of a variable optical attenuator to emulate fibre loss or a real fibre spool. In Bob, an external phase modulator was used to perform active measurement basis switching. The phase modulator added 2.4 dB to the receiver circuit, bringing the total Bob loss to 4.2 dB. The active components of the system were controlled using field programmable gate arrays (FPGAs) adapted from a previously published standalone chip-based QKD system\cite{Paraiso.2021}. 

Qubits were encoded in four states of the equatorial plane of the time-bin Bloch sphere, with a majority basis preferentially selected with probability 0.9375. The three decoy intensity states (signal, decoy, vacuum) had flux ratios 1:0.25:0.01 and occurred with probabilities 14:1:1\cite{Hwang.2003}. The signal flux at the output of Alice was set to 0.4 photons per pulse. A pseudo-random repeating pattern of 4096 pulses was used, with performance analysed in post-processing with finite key size blocks of 5 Mbit \cite{Lucamarini.2013}. We use a security parameter of $10^{-10}$ and an error correction efficiency of 1.15. 

\begin{figure}
    \centering
    \includegraphics[width=0.8\linewidth]{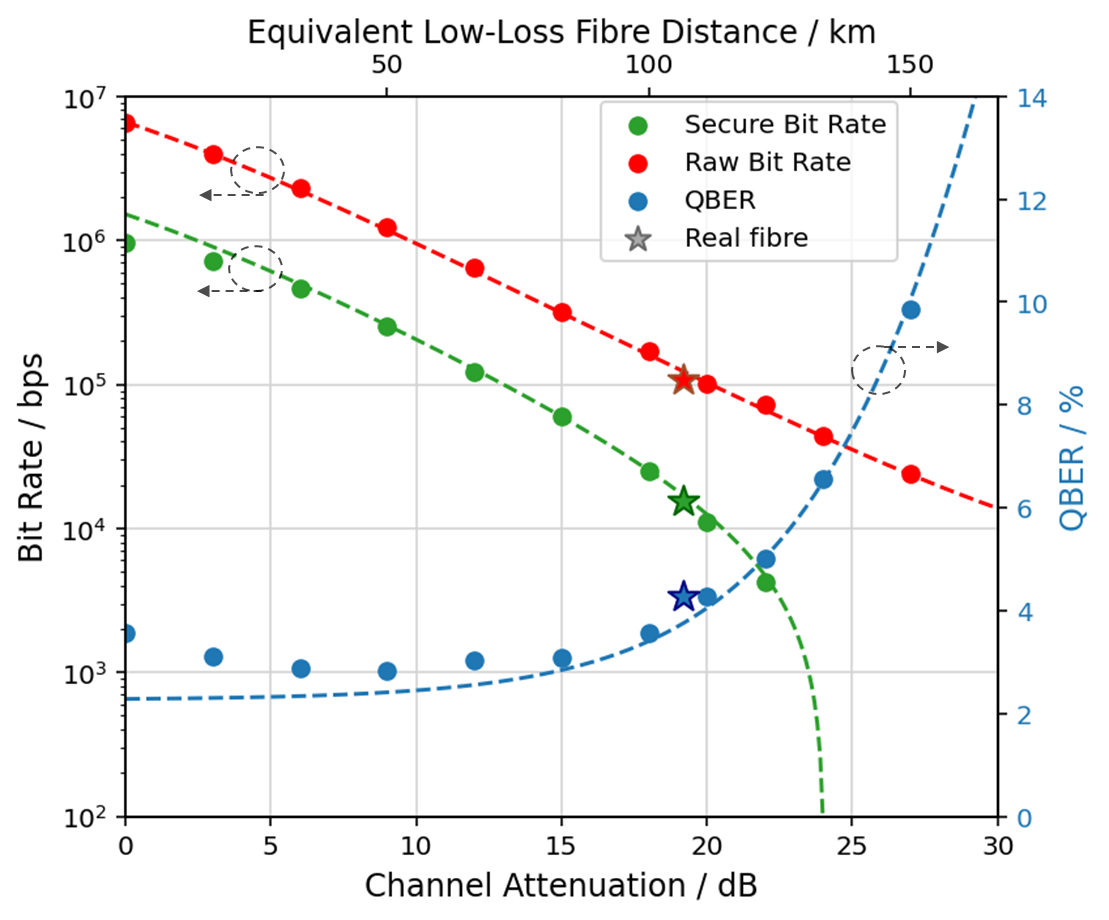}
    \caption{Plot of raw bit rate, QBER and secure bit rate against channel attenuation with detectors at -30{\textdegree}C. The loss of a variable optical attenuator in the quantum channel was gradually increased to emulate increasing fibre channel distance. Equivalent fibre channel distance is calculated at 0.18 dB/km. Data was also taken over a real 100 km fibre spool (stars), with a loss of 19.2 dB (elevated due to fibre connector losses). The dotted lines show the results of simulations based on measured system parameters. }
    \label{fig:QKD_minus30C}
\end{figure}

The results of the QKD experiment are shown in Figure \ref{fig:QKD_minus30C}. We demonstrate positive secure key rates across the channel attenuation range from 0 dB to 22 dB, with a peak key rate of just under 1 Mbps. Furthermore, our integrated-SPAD receiver was capable of producing positive secure key rates at 100 km of real fibre, practical for most metropolitan network applications. The maximum secure key rate at low attenuations was limited by the deadtime-related saturation of the detectors, which limits the raw count rate and elevates the QBER. The minimum QBER of the system was just under 3\%, with the primary error contributions coming from the afterpulsing and dark counts of the detectors.

Whilst  -30{\textdegree}C is an achievable temperature for thermo-electric coolers, it necessitates hermetic packaging of the receiver assembly to prevent condensation. With this in mind, we experimented with room temperature operation of the SPAD array (see Supplemental Document Section 5). Increasing detector temperature increases the dark count rate, which decreases the maximum range of the system. However, the change also decreases afterpulsing probability and marginally increases SPDE \cite{Comandar.2014}. The room-temperature detectors were operated at 19.0\% SPDE, with 1.47\% afterpulsing, 100 ns deadtime and a combined dark count rate of 130 kHz. This decreased the maximum tolerable channel attenuation to 10 dB ($\sim$50 km fibre), though the peak secure key rate at 0 dB increased to 2.1 Mbps. This operating state could prove superior for low-cost links in a short-range access network.

\section{Discussion}

We have demonstrated a QKD receiver with hybrid-integrated non-cryogenic detectors. The results highlight a realistic pathway toward the manufacture of compact QKD receivers with non-cryogenic cooling requirements. It remains the case that the highest performance detectors for QKD are superconducting nanowire single photon detectors (SNSPDs), which provide superior detection efficiency (which increases secure bit rate) and lower dark count rate (which increases range). However, the significant size, weight, power and cost of SNSPDs systems seems sure to limit them to the most cost insensitive of applications. The integrated SPAD-array-based devices demonstrated within this work are best-suited to fulfilling the large-scale deployment of QKD in metropolitan and access quantum networks, where the high-volume production capability and cost efficiency are prioritised over secure bit rate and range. In classical fibre-optic communication, small-form-factor pluggable transceivers, typically only a few centimetres in size, integrate the full photonic stack of lasers, modulators, and detectors. In contrast, quantum networks have relied on external detector systems, precluding similar miniaturisation. Our work suggests a viable route toward achieving this goal. 

In our demonstration, the footprint of Bob is increased by an external phase modulator, used to actively select the measurement basis. However, this is not a fundamental requirement. A fully passive receiver circuit could instead measure orthogonal time-bin bases using a larger waveguide circuit \cite{Tanaka.2012}. This approach requires more detectors, but such scaling is straightforward and cost-effective with the linear SPAD array architecture. A fully-passive receiver would also significantly reduce total optical loss by eliminating the discrete phase modulator, potentially lowering receiver insertion loss to $\sim$2~dB. This would represent one of the lowest publicly disclosed losses for a QKD receiver PIC. Future work may explore this direction.

As with any PIC-based device, packaging is as critical as fabrication, and can easily be the most expensive part of the manufacturing process. A key focus for future development will be addressing the engineering challenges of packaging the hybrid receiver assembly into a compact form factor. This involves permanently bonding the waveguide chip to the SPAD PCB while maintaining optical alignment and encapsulating the assembly in a hermetically sealed environment. While non-trivial, these are familiar problems in the packaging industry. Moreover, the option to operate the detectors at room temperature loosens the requirements for a hermetically sealed enclosure.

Beyond QKD, the ability to produce scalable, waveguide-coupled single-photon detector arrays at telecom wavelengths is compelling. Such arrays could find applications in emerging areas of quantum sensing and communication. Our gated SPAD operation is particularly well-suited to quantum communication, where photon arrival times are well-defined, but the architecture may also benefit sensing applications. Perhaps most notably, the suppression of SPAD array crosstalk through fast gating is a useful technique. Investigation and modelling of the mechanisms of fast-gating crosstalk suppression could yield further concepts for crosstalk reduction algorithms in free-running SPAD arrays\cite{Chau2015}. Taken together, these results highlight a versatile detector platform with broad relevance to future quantum technologies.

\begin{backmatter}

\bmsection{Supplementary Material}
See Supplemental Document for supporting content. 

\bmsection{Acknowledgment}
We thank Robert Woodward for supplying BB84 finite key analysis code and guidance. We thank M. Sanzaro for productive conversations and ideas. J.A.D. thanks R.V. Penty for his guidance and supervision. J.A.D. acknowledges PhD funding from the UK’s Engineering and Physical Sciences Research Council under the Industrial Cooperative Awards in Science \& Technology (CASE) programme. 

\bmsection{Disclosures}
The authors declare no competing interests. 

\bmsection{Data Availability Statement}
Data underlying the results presented in this paper are not publicly available at this time but may be obtained from the authors upon reasonable request. 

\end{backmatter}

\bibliography{export,rose}

\begin{thebibliography}{10}
\newcommand{\enquote}[1]{``#1''}

\bibitem{Honjo.2004}
T.~Honjo, K.~Inoue, and H.~Takahashi, \enquote{Differential-phase-shift quantum key distribution experiment with a planar light-wave circuit mach-zehnder interferometer,} {\protect\JournalTitle{Optics letters}} \textbf{29}, 2797--2799 (2004).

\bibitem{Tanaka.2012}
A.~Tanaka, M.~Fujiwara, K.~ichiro Yoshino, \emph{et~al.}, \enquote{High-speed quantum key distribution system for 1-mbps real-time key generation,} {\protect\JournalTitle{IEEE Journal of Quantum Electronics}} \textbf{48}, 542--550 (2012).

\bibitem{Sax2023}
R.~Sax, A.~Boaron, G.~Boso, \emph{et~al.}, \enquote{High-speed integrated qkd system,} {\protect\JournalTitle{Photonics Research}} \textbf{11}, 1007 (2023).

\bibitem{Sibson.2017}
P.~Sibson, C.~Erven, M.~Godfrey, \emph{et~al.}, \enquote{Chip-based quantum key distribution,} {\protect\JournalTitle{Nature communications}} \textbf{8}, 13984 (2017).

\bibitem{Paraiso.2019}
T.~K. Paraïso, I.~de~Marco, T.~Roger, \emph{et~al.}, \enquote{A modulator-free quantum key distribution transmitter chip,} {\protect\JournalTitle{npj Quantum Information}} \textbf{5}, 145 (2019).

\bibitem{Semenenko.2020}
H.~Semenenko, P.~Sibson, A.~Hart, \emph{et~al.}, \enquote{Chip-based measurement-device-independent quantum key distribution,} {\protect\JournalTitle{Optica}} \textbf{7}, 238 (2020).

\bibitem{Paraiso.2021}
T.~K. Paraïso, T.~Roger, D.~G. Marangon, \emph{et~al.}, \enquote{A photonic integrated quantum secure communication system,} {\protect\JournalTitle{Nature Photonics}} \textbf{15}, 850--856 (2021).

\bibitem{Sibson.2017b}
P.~Sibson, J.~E. Kennard, S.~Stanisic, \emph{et~al.}, \enquote{Integrated silicon photonics for high-speed quantum key distribution,} {\protect\JournalTitle{Optica}} \textbf{4}, 172 (2017).

\bibitem{Ding.2017}
Y.~Ding, D.~Bacco, K.~Dalgaard, \emph{et~al.}, \enquote{High-dimensional quantum key distribution based on multicore fiber using silicon photonic integrated circuits,} {\protect\JournalTitle{npj Quantum Information}} \textbf{3} (2017).

\bibitem{Bunandar.2018}
D.~Bunandar, A.~Lentine, C.~Lee, \emph{et~al.}, \enquote{Metropolitan quantum key distribution with silicon photonics,} {\protect\JournalTitle{Physical Review X}} \textbf{8} (2018).

\bibitem{Zhang.2019}
G.~Zhang, J.~Y. Haw, H.~Cai, \emph{et~al.}, \enquote{An integrated silicon photonic chip platform for continuous-variable quantum key distribution,} {\protect\JournalTitle{Nature Photonics}} \textbf{13}, 839--842 (2019).

\bibitem{Geng.2019}
W.~Geng, C.~Zhang, Y.~Zheng, \emph{et~al.}, \enquote{Stable quantum key distribution using a silicon photonic transceiver,} {\protect\JournalTitle{Optics express}} \textbf{27}, 29045--29054 (2019).

\bibitem{Cao.2020}
L.~Cao, W.~Luo, Y.~X. Wang, \emph{et~al.}, \enquote{Chip-based measurement-device-independent quantum key distribution using integrated silicon photonic systems,} {\protect\JournalTitle{Physical Review Applied}} \textbf{14} (2020).

\bibitem{Wei.2020}
K.~Wei, W.~Li, H.~Tan, \emph{et~al.}, \enquote{High-speed measurement-device-independent quantum key distribution with integrated silicon photonics,} {\protect\JournalTitle{Physical Review X}} \textbf{10} (2020).

\bibitem{Avesani.2021}
M.~Avesani, L.~Calderaro, M.~Schiavon, \emph{et~al.}, \enquote{Full daylight quantum-key-distribution at 1550 nm enabled by integrated silicon photonics,} {\protect\JournalTitle{npj Quantum Information}} \textbf{7} (2021).

\bibitem{Beutel.2021}
F.~Beutel, H.~Gehring, M.~A. Wolff, \emph{et~al.}, \enquote{Detector-integrated on-chip qkd receiver for ghz clock rates,} {\protect\JournalTitle{npj Quantum Information}} \textbf{7} (2021).

\bibitem{Lin2025}
Z.~Lin, Y.~Gao, L.~Zhou, \emph{et~al.}, \enquote{Integrated lithium niobate photonics for high-speed quantum key distribution,} {\protect\JournalTitle{Optica Quantum}} \textbf{3}, 195 (2025).

\bibitem{Dolphin2023}
J.~A. Dolphin, T.~K. Paraïso, H.~Du, \emph{et~al.}, \enquote{A hybrid integrated quantum key distribution transceiver chip,} {\protect\JournalTitle{npj Quantum Information}} \textbf{9}, 84 (2023).

\bibitem{Ceccarelli.2021}
F.~Ceccarelli, G.~Acconcia, A.~Gulinatti, \emph{et~al.}, \enquote{Recent advances and future perspectives of single-photon avalanche diodes for quantum photonics applications,} {\protect\JournalTitle{Advanced Quantum Technologies}} \textbf{4} (2021).

\bibitem{Zheng.2021}
X.~Zheng, P.~Zhang, R.~Ge, \emph{et~al.}, \enquote{Heterogeneously integrated, superconducting silicon-photonic platform for measurementdevice-independent quantum key distribution,} {\protect\JournalTitle{Advanced Photonics}} \textbf{3} (2021).

\bibitem{Martinez.2017}
N.~J.~D. Martinez, M.~Gehl, C.~T. Derose, \emph{et~al.}, \enquote{Single photon detection in a waveguide-coupled ge-on-si lateral avalanche photodiode,} {\protect\JournalTitle{Optics Express}} \textbf{25}, 16130 (2017).

\bibitem{Wang.2023}
H.~Wang, Y.~Shi, Y.~Zuo, \emph{et~al.}, \enquote{High-performance waveguide coupled germanium-on-silicon single-photon avalanche diode with independently controllable absorption and multiplication,} {\protect\JournalTitle{Nanophotonics}} \textbf{12} (2023).

\bibitem{Zhang2023}
J.~Zhang, H.~Xu, G.~Zhang, \emph{et~al.}, \enquote{Hybrid and heterogeneous photonic integrated near-infrared ingaas/inalas single-photon avalanche diode,} {\protect\JournalTitle{Quantum Science and Technology}} \textbf{8} (2023).

\bibitem{Alimi2025}
Y.~Alimi, B.~Guilhabert, D.~Jevtics, \emph{et~al.}, \enquote{Micro-transfer printing of ingaas/inp avalanche photodiode on si substrate,} {\protect\JournalTitle{Semiconductor Science and Technology}} \textbf{40} (2025).

\bibitem{Ren2024}
X.~Ren, D.~Liu, M.~Hu, \emph{et~al.}, \enquote{Hybrid integration of ingaas/inp single photon avalanche diodes array and silicon photonics chip,} in \emph{Opto-Electronics and Communications Conference, OECC,}  (Institute of Electrical and Electronics Engineers Inc., 2024).

\bibitem{Han2020}
S.~Han, J.~Park, S.~Yoo, and K.~Yu, \enquote{Lateral silicon photonic grating-to-fiber coupling with angle-polished silica waveguide blocks,} {\protect\JournalTitle{Optics Express}} \textbf{28}, 8811 (2020).

\bibitem{pellegrini2006design}
S.~Pellegrini, R.~E. Warburton, L.~J. Tan, \emph{et~al.}, \enquote{Design and performance of an ingaas-inp single-photon avalanche diode detector,} {\protect\JournalTitle{IEEE Journal of Quantum Electronics}} \textbf{42}, 397--403 (2006).

\bibitem{Jiang2007}
X.~Jiang, M.~A. Itzler, R.~Ben-Michael, and K.~Slomkowski, \enquote{Ingaasp-inp avalanche photodiodes for single photon detection,} {\protect\JournalTitle{IEEE Journal on Selected Topics in Quantum Electronics}} \textbf{13}, 895--904 (2007).

\bibitem{ramirez2009dependence}
D.~A. Ramirez, M.~M. Hayat, and M.~A. Itzler, \enquote{Dependence of the performance of single photon avalanche diodes on the multiplication region width,} {\protect\JournalTitle{IEEE Journal of Quantum Electronics}} \textbf{44}, 1188--1195 (2009).

\bibitem{chen2023guard}
Y.-C. Chen, R.-H. Yan, H.-C. Huang, \emph{et~al.}, \enquote{Guard ring design to prevent edge breakdown in double-diffused planar ingaas/inp avalanche photodiodes,} {\protect\JournalTitle{Materials}} \textbf{16}, 1667 (2023).

\bibitem{yuan2007high}
Z.~Yuan, B.~Kardynal, A.~Sharpe, and A.~Shields, \enquote{High speed single photon detection in the near infrared,} {\protect\JournalTitle{Applied Physics Letters}} \textbf{91} (2007).

\bibitem{Yuan2008}
Z.~L. Yuan, A.~R. Dixon, J.~F. Dynes, \emph{et~al.}, \enquote{Gigahertz quantum key distribution with ingaas avalanche photodiodes,} {\protect\JournalTitle{Applied Physics Letters}} \textbf{92} (2008).

\bibitem{Yuan.2018}
Z.~Yuan, A.~Murakami, M.~Kujiraoka, \emph{et~al.}, \enquote{10-mb/s quantum key distribution,} {\protect\JournalTitle{Journal of Lightwave Technology}} \textbf{36}, 3427--3433 (2018).

\bibitem{Makarov2006}
V.~Makarov, A.~Anisimov, and J.~Skaar, \enquote{Effects of detector efficiency mismatch on security of quantum cryptosystems,} {\protect\JournalTitle{Physical Review A}} \textbf{74} (2006).

\bibitem{Fung.2009}
C.-H.~F. Fung, K.~Tamaki, B.~Qi, \emph{et~al.}, \enquote{Security proof of quantum key distribution with detection efficiency mismatch,} {\protect\JournalTitle{Quantum Info. Comput.}} \textbf{9}, 131--165 (2009).

\bibitem{Itzler2010}
M.~A. Itzler, M.~Entwistle, M.~Owens, \emph{et~al.}, \enquote{Geiger-mode avalanche photodiode focal plane arrays for three-dimensional imaging ladar,} in \emph{Infrared Remote Sensing and Instrumentation XVIII,}  vol. 7808 (SPIE, 2010), p. 78080C.

\bibitem{Calandri2016}
N.~Calandri, M.~Sanzaro, L.~Motta, \emph{et~al.}, \enquote{Optical crosstalk in ingaas/inp spad array: Analysis and reduction with fib-etched trenches,} {\protect\JournalTitle{IEEE Photonics Technology Letters}} \textbf{28}, 1767--1770 (2016).

\bibitem{Liu2022}
C.~Liu, H.~F. Ye, and Y.~L. Shi, \enquote{Advances in near-infrared avalanche diode single-photon detectors,} {\protect\JournalTitle{Chip}} \textbf{1} (2022).

\bibitem{Pesantes.2023}
K.~A.~H. Pesantes, E.~Conca, S.~Riccardo, \emph{et~al.}, \enquote{8-channel time-of-flight single-photon detection module based on ingaas/inp spad array,} in \emph{PRIME 2023 - 18th International Conference on Ph.D Research in Microelectronics and Electronics, Proceedings,}  (Institute of Electrical and Electronics Engineers Inc., 2023), pp. 93--96.

\bibitem{Tang.2024}
Y.~Tang, R.~Wang, X.~Yang, \emph{et~al.}, \enquote{High crosstalk suppression in ingaas/inp single-photon avalanche diode arrays by carrier extraction structure,} {\protect\JournalTitle{Nature Communications}} \textbf{15} (2024).

\bibitem{Amer.2021}
O.~Amer, V.~Garg, and W.~O. Krawec, \enquote{An introduction to practical quantum key distribution,}  (2021).

\bibitem{Lucamarini.2013}
M.~Lucamarini, K.~A. Patel, J.~F. Dynes, \emph{et~al.}, \enquote{Efficient decoy-state quantum key distribution with quantified security,} {\protect\JournalTitle{Optics express}} \textbf{21}, 24550--24565 (2013).

\bibitem{Itzler2009}
M.~A. Itzler, M.~Entwistle, M.~Owens, \emph{et~al.}, \enquote{Inp-based geiger-mode avalanche photodiode arrays for three-dimensional imaging at 1.06 um,} in \emph{Advanced Photon Counting Techniques III,}  vol. 7320 (SPIE, 2009), p. 73200O.

\bibitem{Vila2014}
A.~Vila, E.~Vilella, O.~Alonso, and A.~Dieguez, \enquote{Crosstalk-free single photon avalanche photodiodes located in a shared well,} {\protect\JournalTitle{IEEE Electron Device Letters}} \textbf{35}, 99--101 (2014).

\bibitem{Zhang2019}
H.~Zhang, L.~Su, D.~Zhou, \emph{et~al.}, \enquote{Crosstalk analysis of sic ultraviolet single photon avalanche photodiode arrays,} {\protect\JournalTitle{IEEE Photonics Journal}} \textbf{11} (2019).

\bibitem{Broquin2021}
J.~E. Broquin and S.~Honkanen, \enquote{Integrated photonics on glass: A review of the ion-exchange technology achievements,}  (2021).

\bibitem{Brusberg2025}
L.~Brusberg, M.~Granados-Baez, R.~D. Schilling, \emph{et~al.}, \enquote{Optical design and applications for ion-exchanged glass waveguide circuits,} {\protect\JournalTitle{IEEE Transactions on Components, Packaging and Manufacturing Technology}} \textbf{15}, 1614--1624 (2025).

\bibitem{Hwang.2003}
W.-Y. Hwang, \enquote{Quantum key distribution with high loss: toward global secure communication,} {\protect\JournalTitle{Physical review letters}} \textbf{91}, 57901 (2003).

\bibitem{Comandar.2014}
L.~C. Comandar, B.~Fröhlich, M.~Lucamarini, \emph{et~al.}, \enquote{Room temperature single-photon detectors for high bit rate quantum key distribution,} {\protect\JournalTitle{Applied Physics Letters}} \textbf{104} (2014).

\bibitem{Chau2015}
Q.~Chau, X.~Jiang, M.~A. Itzler, \emph{et~al.}, \enquote{Analysis and modeling of optical crosstalk in inp-based geiger-mode avalanche photodiode fpas,} in \emph{Advanced Photon Counting Techniques IX,}  vol. 9492 (SPIE, 2015), p. 94920O.

\end{thebibliography}

\end{document}